\documentclass[conference]{IEEEtran}
\IEEEoverridecommandlockouts

\usepackage{cite}
\usepackage{amsmath,amssymb,amsfonts}
\usepackage{algorithmic}
\usepackage{graphicx}
\usepackage{textcomp}
\usepackage{xcolor}

\usepackage[utf8]{inputenc}
\usepackage{subfig}
\usepackage[T1]{fontenc}
\usepackage[hidelinks]{hyperref}
\usepackage{wrapfig}
\usepackage{dirtree}

\usepackage{todonotes}

\usepackage{changes}
\definechangesauthor[name=nicolas, color=teal]{nr}
\definechangesauthor[name=fabien, color=red]{fg}
\definechangesauthor[name=maxime, color=blue]{cf}

\def\BibTeX{{\rm B\kern-.05em{\sc i\kern-.025em b}\kern-.08em
    T\kern-.1667em\lower.7ex\hbox{E}\kern-.125emX}}
\begin{document}

\title{OLIVAW: ACIMOV's GitHub robot assisting agile collaborative ontology development}

\author{\IEEEauthorblockN{Nicolas Robert}
\IEEEauthorblockA{\href{https://inria.fr/fr}{Inria}, \href{https://univ-cotedazur.fr/}{UNICA}, \textit{\href{https://www.cnrs.fr/fr}{CNRS}, \href{https://www.i3s.unice.fr/fr/}{I3S}}\\
Biot, France \\
\href{https://orcid.org/0009-0009-2595-6168}{0009-0009-2595-6168}}
\and
\IEEEauthorblockN{Fabien Gandon}
\IEEEauthorblockA{\href{https://inria.fr/fr}{Inria}, \href{https://univ-cotedazur.fr/}{UNICA}, \textit{\href{https://www.cnrs.fr/fr}{CNRS}, \href{https://www.i3s.unice.fr/fr/}{I3S}}\\
Biot, France \\
\href{https://orcid.org/0000-0003-0543-1232}{0000-0003-0543-1232}}
\and
\IEEEauthorblockN{Maxime Lefrançois}
\IEEEauthorblockA{Mines Saint-Etienne, Univ Clermont Auvergne,\\ INP Clermont Auvergne, CNRS, UMR 6158 LIMOS,\\ Saint-\'Etienne, France \\
\href{https://orcid.org/0000-0001-9814-8991}{0000-0001-9814-8991}}
}

\maketitle

\begin{abstract}
Agile and collaborative approaches to ontologies design are crucial because they contribute to making them user-driven, up-to-date, and able to evolve alongside the systems they support, hence proper continuous validation tooling is required to ensure ontologies match developers' requirements 
all along their development. We propose OLIVAW (Ontology Long-lived Integration Via ACIMOV Workflow), a tool supporting the ACIMOV methodology on GitHub. It relies on W3C Standards to assist the development of modular ontologies through GitHub Composite Actions, pre-commit hooks, or a command line interface. OLIVAW was tested on several ontology projects to ensure its usefulness, genericity and reusability. A template repository is available for a quick start. OLIVAW is published under the LGPL-2.1 license and archived on Software Heritage and Zenodo.
\end{abstract}


\section{Introduction}\label{sec:intro}
Ontologies, schemata, and vocabularies, are more and more used in information systems.
They provide a structured way to define, organize, and link knowledge, enabling more effective data and application integration. 
They act as an underlying reference for knowledge representation, allowing information systems to achieve semantic interoperability in increasingly complex and distributed digital ecosystems.


In that context, agile and collaborative approaches to ontology development are crucial to make ontologies user-driven, consensual, up-to-date, and able to evolve alongside the systems they support. Mirroring the best practices of modern software development, ontology development benefits from being iterative, responsive to feedback, and collaborative
, ultimately resulting in ontologies that are more resilient, useful, and aligned with user needs~\cite{hannou:hal-04187236}.

The distributed version management software \href{https://git-scm.com/}{Git} is widely adopted in collaborative software development.
Git platforms such as \href{https://github.com/}{GitHub} or \href{https://about.gitlab.com/}{GitLab} have popularised development workflows centred on branches, forks, and pull/merge requests, support agile project management with issues and milestones, and support DevOps principles through continuous integration and deployment. These continuous integration mechanisms are ensured by \emph{GitHub Actions} on Github and \emph{GitLab CI/CD} on GitLab. GitHub also provides the \emph{GitHub Gist} service allowing to share text snippets referred to as \emph{gist files}.


Leveraging agility and DevOps principles with a standard git-based approach that is relying on the CI/CD mechanisms is a distinguishing feature of the ACIMOV ontology engineering methodology (Agile and Continuous Integration for Modular Ontologies and Vocabularies)~\cite{hannou:hal-04187236}. 
For example, the issue tracker can be used to discuss and document ontology engineering decisions. Labels and milestones can be used to categorise and prioritise requirements and manage the backlog. Continuous integration and deployment workflows can perform syntactic and semantic checks on the ontology, and generate and publish its documentation.


This paper presents OLIVAW (Ontology Long-lived Integration Via ACIMOV Workflow), a tool that supports the ACIMOV methodology on GitHub to assist agile collaborative ontology development. OLIVAW relies on W3C Standards, and is usable as a standalone library through a command line interface, as pre-commit hooks, or as GitHub Actions. OLIVAW focuses on continuous integration and can be combined with other existing software for other aspects of development such as documentation or deployment.
OLIVAW is \emph{FAIR} since its (1) \emph{Findable}  by its Zenodo DOI, Software Heritage SWHID and HAL-ID; (2) \emph{Accessible} by Zenodo, Software Heritage, GitHub, Pipy ; (3) \emph{Interoperable} by its use of development standards such as GitHub Actions, PyPI, W3C standards ; and (4) \emph{Reusable} by its LGPL license, its documentation and repository template.

The next sections describe \href{https://github.com/Wimmics/olivaw}{OLIVAW} starting with the related work (Section \ref{sec:related-work}) and terminology (Section \ref{sec:terminology}). We then introduce the general principles of OLIVAW and each family of tests it can perform and reports it produces (Section \ref{sec:OLIVAW}). We explain the different alternatives to deploy OLIVAW (Section \ref{sec:deploy}) and report on our experimentations and evaluations (Section \ref{sec:experimentation}), before concluding.

\section{Related Work}
\label{sec:related-work}

Several ontology engineering methodologies are inspired by the principles of agile software engineering, which promote collaboration between developers and stakeholders by producing regular updates of the product. Among these methods, AMOD~\cite{abdelghany2019agile} and CD-OAM \cite{takhom2020collaborative} are based on SCRUM. XPOD~\cite{conf/fomi/ShariflooS08}, eXtreme ontology method~\cite{hristozova2002extreme}, and eXtreme Design (XD)~\cite{de2013organizing} are based on eXtreme Programming. The Lean Ontology Development (LOD) \cite{cummings2018lean} is inspired by the Lean approach: Build-Measure-Learn. SAMOD \cite{peroni2016samod} is revisiting the motivating scenarios and competency questions of Uschold and Gruninger~\cite{uschold_gruninger_1996}, additionally considering ontology modules and test-driven development. 
Some tools exist to support some of these methodologies. For example, XDTesting \cite{ciroku2023supporting} is an automated test tool supporting the XD methodology. Three types of tests are available, namely the Competency Question Verification which consists in testing a competency question over a dataset, an Inference Verification which consists in checking if the entailment creates some required triples, and the Error Provocation which consists in feeding a modelet with a dataset that is incomplete or incorrect in order to check how the entailment behaves. 

Just as agility aims to improve collaborations between software project customers and developers, DevOps improves collaborations between developers and IT operations professionals. 
Jenkins, Travis CI, Circle CI, GitLab CI/CD, Github Actions, are all frameworks that allow to specify CI/CD workflows that will be executed automatically when, for example, a commit is pushed to the server. Before even these frameworks existed, VoCol \cite{halilaj2016vocol} and OnToology \cite{alobaid2019automating} supported CI/CD in ontology engineering using Github applications\footnote{{\url{https://docs.github.com/en/developers/apps}}}. 
The \emph{Ontology Development Kit} (ODK) \cite{matentzoglunicolas} uses Travis CI to run CI/CD workflows with the ROBOT tool~\cite{jackson2019robot} developed by the Open Biological and Biomedical Ontologies (OBO) community. 
Different ontology projects use CI/CD workflows, such as the Financial Industry Business Ontology (FIBO)~\cite{neuhaus2022infrastructure}, the International Data Spaces Information Model (IDSA)~\cite{bader2020international}, the CASE Cyber Ontology\footnote{\url{https://github.com/marketplace/actions/case-ontology-validator}} and the Smart Applications REFerence Ontology (SAREF)~\cite{saref10154112}. The latter uses a dedicated development methodology and a dedicated software, the SAREF-Pipeline~\cite{lefranccois2023saref}, that is used to verify the quality of contributions and generate the documentation website. To set up CI/CD workflows for new ontology projects, some Github actions are available on the Github marketplace for running RDFLint\footnote{\url{https://github.com/marketplace/actions/setup-rdflint}}, validating RDF syntaxes\footnote{\url{https://github.com/marketplace/actions/rdf-syntax-check}}\textsuperscript{,}\footnote{\url{https://github.com/marketplace/actions/validate-rdf-with-jena}}, or validating RDF files against SHACL shapes\footnote{\url{https://github.com/marketplace/actions/validate-shacl}} or ShEx \cite{publio2022ontolo}.


\begin{figure}
    \centering
    \includegraphics[width=0.99\linewidth]{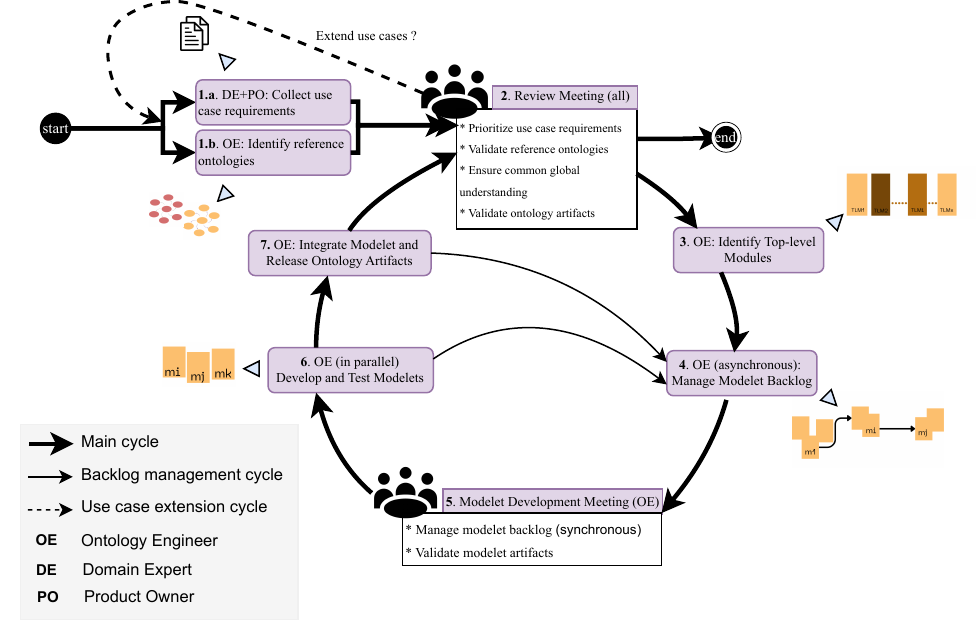}
    \caption{Overview of the ACIMOV methodology~\cite{hannou:hal-04187236}}
    \label{fig:acimov-methodo}
\end{figure}

ACIMOV~\cite{hannou:hal-04187236} extends SAMOD and explicitly adopts the standard git-based approach for coding, leveraging agility and DevOps principles. It addresses modularity needs with ontologies being designed as a set of stand-alone modules called \emph{modelets}, describing specific aspects and covering focused sets of requirements~\cite{hannou:hal-04187236}. The assignment and the backlog of modelets is done through the issue tracker. ACIMOV emphasizes collaboration among team members, stakeholders, and domain experts throughout the ontology development process. Each ontology engineer contributing to the repository can take the charge of developing and testing a modelet's artifacts, typically including: motivating scenarios, competency questions, specification of classes, properties and individuals, some RDF test data for a use case, SPARQL queries representing competency questions to be tested over test data, etc.
Following ACIMOV, modelets require \emph{testing} to check the quality and coherence of the ontological description, \emph{validation} of successful testing results and a decision from the ontology engineers for integration to a new version of the ontology and a release. In other words, there is a need for tooling. ACIMOV is pure methodology with no tooling and the OLIVAW framework introduced in this article is meant to provide such tooling, supporting continuous integration for ACIMOV. It covers a range of tests including Competency Question Verification and Inference Verification, and provides even more e.g. ontology profile diagnoses.

In the next section we recall some RDF vocabularies and core concepts from SAMOD and ACIMOV that are needed to understand OLIVAW.

\section{Terminology}\label{sec:terminology}

OLIVAW is grounded in the Semantic Web stack using RDF~\cite{Lanthaler:14:RCA}, OWL~\cite{Parsia:12:OWO}, SPARQL~\cite{Seaborne:13:SQL} and SHACL~\cite{Kontokostas:17:SCL}. It also uses PROV-O~\cite{Sahoo:13:PTP}, and the Evaluation and Report Language vocabulary (EARL)~\cite{Abou-Zahra:17:ERL}. 

EARL is an RDF vocabulary allowing the representation of test reports. Each test report entry is seen as \textbf{Assertion}, which is an aggregation of : an \textbf{Assertor} (the actor who ran the test), a \textbf{TestSubject} (the tested resource, which can be composed of several files) a \textbf{TestCriterion} (the constraint to be tested against the TestSubject) and a \textbf{TestResult} (the result of the test that was run, which is an aggregation of Outcomes that provide precise information).
%
%
A TestResult is composed of outcomes that can have for output type: \textbf{Pass} if the subject passed the test; \textbf{NotTested} if the test could not be run because some prerequisite statements were not validated (example: for a car, test ``gear shift'' could not be run because test "turn on" was not validated); \textbf{CannotTell} if the automated test detected some unexpected behavior where a human is required to conclude if this is an error; and  \textbf{Fail} if the subject failed the test.



SAMOD grounds the development of ontologies on \textbf{Motivating Scenarios}, which are practical descriptions of situations related to a domain. A motivating scenario is used to justify the introduction of new terms in some iteration of the ontology development. 
\textbf{Modelets} are stand-alone models that formalize the new terms of a motivating scenario.
Requirements are first expressed informally using \textbf{Competency Question} in natural language, then are implemented in SPARQL or any other formal language.
\textbf{Dataset} (ABox in SAMOD) are small data graphs that model part of the motivating scenario, and are used to validate modelets against formalized competency questions. 
The final ontology is the union of all modelets.
ACIMOV extends SAMOD to support the development of modular and multi-domain ontologies. 
Motivating scenarios are grouped in domains. Modules are ontology fragments that merge an arbitrary set of modelets.
\textbf{Use-cases} are larger data graphs that illustrate all of the motivating scenarios related to a specific domain.
As OLIVAW supports ACIMOV, three types of tests will be involved depending on the type of resources they target: (1) \textbf{Model tests} for modelets and modules, which we refer to as \textbf{ontology fragments}; \textbf{Data tests} for datasets and use-cases, which we refer to as \textbf{data fragments}; and \textbf{Query tests} for formal competency questions.

\section{OLIVAW Framework and Dev. Workflow}\label{sec:OLIVAW}

OLIVAW is an open source python framework that supports ACIMOV, published under the \href{https://spdx.org/licenses/LGPL-2.1}{LGPL-2.1 license}. 
It is available as \href{https://github.com/Wimmics/olivaw}{a GitHub repository}\footnote{\url{https://github.com/Wimmics/olivaw}}, \href{https://pypi.org/project/olivaw/}{a PyPI package}\footnote{\url{https://pypi.org/project/olivaw/}}, \href{https://archive.softwareheritage.org/browse/origin/directory/?origin_url=https://github.com/Wimmics/olivaw}{a Software Heritage Archive}\footnote{\url{https://archive.softwareheritage.org/browse/origin/directory/?origin_url=https://github.com/Wimmics/olivaw}} and \href{https://doi.org/10.5281/zenodo.14288084}{a Zenodo record}\footnote{\url{https://doi.org/10.5281/zenodo.14288084}}. It has a full \href{https://github.com/Wimmics/olivaw/tree/main/docs}{online documentation with examples}\footnote{\url{https://github.com/Wimmics/olivaw/tree/main/docs}}. It is documented using Sphinx docstrings, function signatures using \texttt{typing} and many \texttt{README.md} files for optimal code readability, maintainability and extensibility. The documentation also provides OLIVAW reports examples for several ontologies.\footnote{\url{https://github.com/Wimmics/olivaw/tree/main/docs/examples}}

\subsection{OLIVAW ontology project structure}\label{sec:olivaw-project-structure}

\begin{figure*}
\centering
\begin{minipage}[b]{.45\textwidth}
\begin{scriptsize}
\dirtree{%
.1 ./.
.1 .acimov/.
.2 parameters.json.
.2 output/.
.2 custom-tests/.
.3 model/.
.3 data/.
.1 .github/.
.2 workflows/ \# for ci/cd.
.1 domains/.
.2 domain-1/.
.3 motivating-scenario-1/.
.4 onto.ttl.
.4 dataset.ttl.
.4 q1.rq.
.1 src/.
.2 module.ttl.
.1 use-cases/.
.2 domain-1/.
.3 dataset.ttl.
}\end{scriptsize}
  \caption{Repository structure}
  \label{fig:repo-structure}
\end{minipage}%
\begin{minipage}[b]{.45\textwidth}
  \centering
    \includegraphics[width=0.95\linewidth]{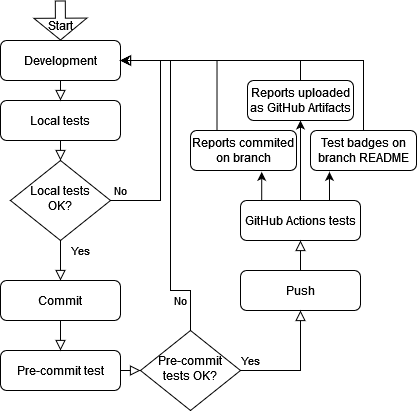}
    \caption{Three complementary executions of tests by OLIVAW}
    \label{fig:olivaw-lifecycle-structure}
\end{minipage}%
\end{figure*}

OLIVAW can support any ontology project matching the minimal folder structure of Figure~\ref{fig:repo-structure}. 
The \href{https://github.com/Wimmics/Olivaw-Template}{OLIVAW repository template}\footnote{\url{https://github.com/Wimmics/Olivaw-Template}} offers a quick start for new ontology projects.
The \texttt{src/} folder contains the ontology modules and \texttt{domains/} contains a sub-folder for each domain. Each domain folder contains a subfolder for each motivating scenario. Each motivating scenario\footnote{Example: \url{https://github.com/HyperAgents/hmas/tree/main/domains/logistics/create-organization}} contains a modelet, a dataset and related competency questions implementations. The \texttt{use-cases/} folder contains a sub-folder for each use case. Each use case may contain one or more data fragments. The \texttt{.acimov/} folder contains files related to the ACIMOV methodology, support tools, and a \texttt{parameters.json} file setting the specific parameters of the project (see Section \ref{sec:deploy}). Files generated by OLIVAW go in \texttt{output/}. Finally, \texttt{custom-tests/} folder is meant to store custom tests to be run against the project (see Section~\ref{sec:custom-tests}).

\subsection{OLIVAW General Overview}\label{sec:overview}


OLIVAW builds on CORESE (Conceptual Resource Search Engine)~\cite{ceres:hal-04170333v1}, which is an open-source semantic web framework\footnote{\url{https://github.com/Wimmics/corese/}} implementing RDF, SPARQL, SHACL, RDFS and OWL RL.
As shown on Figure~\ref{fig:olivaw-lifecycle-structure}, OLIVAW can execute tests at three different and complementary moments: (i) on the computer of the ontology developer, using the command-line actions for some project analysis.     
(ii) as a pre-commit hook, that will check all the files staged for commit and prevent blocking errors. (iii) As a job configured to run during a GitHub Actions CI/CD workflow, for example to analyze the project and generate reports after a push to some branch, or the creation of a pull request. 
Each mechanism can be installed and configured independently.
OLIVAW covers Model tests (see Section~\ref{sec:model-tests}), Data tests (see Section~\ref{sec:data-tests}), and Query tests (see Section \ref{sec:query-tests}). Custom tests are explained in Section~\ref{sec:custom-tests}. %
Tests generate reports in Turtle (see Section~\ref{sec:turtle-report-format}) and Markdown (see Section~\ref{sec:markdown-report-format}).
%
%
These reports are saved in the \texttt{.acimov/output/} folder, on the local drive for tests run locally or on the server for tests run by GitHub Actions, supporting actual branch state control and tracking over time.

\sloppy
During a GitHub Action, dynamic badges are updated on the branch README.md file to provide some basic metrics about how many outcomes are to be found on this project for each outcome type (as defined in Section \ref{sec:terminology}), and also the ontology (merged modules) compatibility w.r.t. each OWL profile (EL/QL/RL). Each of these badges is clickable and points to its related report. 
OLIVAW can follow specific settings to adapt to some particular project needs, such as errors to consider as blocking or tests to discard, through a project file \texttt{.acimov/parameters.json} (see Section \ref{sec:deploy}).





The \href{https://ns.inria.fr/olivaw/}{OLIVAW ontology\footnote{\url{https://ns.inria.fr/olivaw/}}} instanciates some EARL TestCriterion and extends the EARL Fail class to capture the error severity (\texttt{MinorFail} and \texttt{MajorFail} classes). Any error is considered as \texttt{MinorFail} unless it appears in the project list of blocking errors to be considered as a \texttt{MajorFail}. This ontology also includes some terms used in the generated reports e.g.  \texttt{VersionedEntity}, is a class of entities used in the OLIVAW tests, typing the different versions of an entity.

\subsection{Model tests}\label{sec:model-tests}

Model tests cover different levels of checking, from low-level syntax validation to matching model expectations and best practices. These tests are applied on: (1) each module\footnote{Module example from hMAS: \url{https://github.com/HyperAgents/hmas/blob/main/src/regulation.ttl}} individually, (2) each modelet\footnote{Modelet example from hMAS: \url{https://github.com/HyperAgents/hmas/blob/main/domains/manufacturing-environments/discover-signifiers/onto.ttl}} individually, (3) each module merged with related terms from a given modelet, (4) the merge of all the syntactically correct modules and (5) all the syntactically correct modules and modelets merged together (see Section \ref{sec:terminology} for terminology).
All sources of errors are covered on the module level, at the integration stage when merging the terms of a modelet in a module or at the whole ontology level by merging all the ontology modules and modelets together.
Additional custom tests can be added to the model tests for a specific project (see Section \ref{sec:custom-tests}).

\sloppy
OLIVAW includes the following tests by default.
\textbf{RDF syntax validity}: the test subject is loaded in CORESE and should not return any syntax error.
\textbf{Terms are linked to their ontology}: this test consists in checking if for each ontology term, there exists a \texttt{rdfs:isDefinedBy} property with a value for this term. 
\textbf{Domain or range out of vocabulary}: this test consists in checking if each triple with a subject namespaced in the vocabulary and predicate \texttt{rdfs:domain} or \texttt{rdfs:range} has an object namespaced in the ontology or points to an external URI. 
\textbf{Proper use of subset properties}: this test checks if any \texttt{rdfs:subClassOf} predicate has a subject or object that is of type \texttt{rdf:Property} 
 and if any \texttt{rdfs:subPropertyOf} predicate has a subject or object that is of type \texttt{rdfs:Class}. 
\textbf{Differentiation of terms}: checks if any pair of terms has a Levenshtein distance under a given threshold defined in the configuration file (see Section \ref{sec:deploy}), for instance properties \texttt{hasName} and \texttt{has\_name}. 
\textbf{English labels are provided}: this test checks if all the ontology terms have an \texttt{rdfs:label} with a literal value tagged as English (\texttt{@en}). 
\textbf{OWL 2 RL consistency}: this checks the logical consistency of the model within that profile, but not the compatibility with that profile.\footnote{e.g. an ontology containing a \texttt{owl:disjointUnionOf} could be found consistent by an OWL 2 RL reasoner but  would not be compatible with that OWL profile}
\textbf{OWL profile compatibility}: this checks if the tested resource is compatible with the OWL profiles \textit{RL}, \textit{QL} and \textit{EL}.


\subsection{Data tests}\label{sec:data-tests}

Data tests check if the data fragments using the vocabulary respect some standards. These tests are applied on each dataset\footnote{Dataset example from hMAS: \url{https://github.com/HyperAgents/hmas/blob/main/domains/manufacturing-environments/discover-organization/dataset.ttl}} individually and in each use-case\footnote{Use-case example from hMAS: \url{https://github.com/HyperAgents/hmas/blob/182-uc-manufacturing/use-cases/manufacturing/cup-production.ttl}} fragment individually (see Section \ref{sec:terminology}). 
These tests include by default the following tests.
\textbf{Syntax validity}, equivalent to test in Section \ref{sec:model-tests}.
\textbf{OWL RL consistency test}, equivalent to the test in Section \ref{sec:model-tests}.
\textbf{Known Terms}: The test checks if any test subject term belonging to the ontology namespace is not defined in the ontology.
\textbf{Namespace test}: The test checks if any test subject namespace has a Levenshtein distance equal to 1 or 2 from a namespace found on \href{https://prefix.cc}{prefix.cc}, or from another project namespace. This allows the detection of potential typos such as the use of http/https or trailing hash/slash, and is reported as \texttt{CannotTell} since they could be correct yet unknown to OLIVAW.


\subsection{Query tests}\label{sec:query-tests}
Query tests check if the competency questions 
 are matching the expectations and best practices. 
These tests 
include: \textbf{Syntax tests} during which the construction of a SPARQL query AST (abstract syntax tree) is attempted by CORESE, checking the validity of the query syntax. \textbf{Query type test} : any query should use either the \texttt{SELECT} or \texttt{ASK} clause to match an open or closed question in natural language. \textbf{URI validity test}: any URI in the query should conform to the RFC 3986~\cite{rfc3986}. \textbf{Namespace test}: equivalent to Section \ref{sec:data-tests}.

\subsection{Extension points and Custom tests}\label{sec:custom-tests}

Any ontology development project may have additional requirements specific to an application, a community, a domain, an ontology management platform, a methodology, an organization, etc. 
Any project can integrate OLIVAW even if not following the ACIMOV architecture. It is possible to configure OLIVAW to fit a non-ACIMOV architectured project (see \href{https://github.com/Wimmics/olivaw/blob/main/docs/parameters.md}{OLIVAW parameters documentation} for more details). Moreover, SKOS(-XL) thesauri or other vocabularies may require specific tests that OLIVAW does not cover by default but can accommodate.
In order to meet these requirements, OLIVAW supports the implementation of custom tests that can be added to model tests or data tests.

These tests can be implemented as graphs of SHACL shapes that will be tested against the fragments. This feature is meant to enable custom model or data test implementation before ontology or data development, which is a commonly adopted principle in software development (e.g. in V model methodology or in Test Driven Development). This feature also allows custom test reusability from one project to another without any change required on the test itself.

However SHACL can lack the required expressivity for some tests. For example, implementing the detection of a cycle in the hierarchy of properties cannot be done in standard SHACL. Moreover the basic use case of SHACL is to validate a dataset against shapes, targeting some nodes based on classes, URIs or properties that are already known in advance. However none of these targets are known in advance in an ontology development context, which complicates the proper node targeting for upstream test implementation.

In order to make these custom tests powerful and convenient, it is possible to take advantage of all the non-standard SHACL features in CORESE. When additional expressivity is needed, CORESE's implementation of SHACL-SPARQL\footnote{\url{https://www.w3.org/TR/shacl/\#shacl-sparql}} can be used. It brings in SHACL all the expressivity available in SPARQL. CORESE also allows to target triples instead of nodes and brings many new options to implement powerful SHACL paths in order to target and find the fine-grain patterns that are expected to be detected in any specific ontology development project.
This feature is fully documented and several examples can be found in \href{https://github.com/Wimmics/olivaw/blob/main/docs/custom-tests.md}{OLIVAW custom tests documentation and examples}\footnote{\url{https://github.com/Wimmics/olivaw/blob/main/docs/custom-tests.md}}.

\subsection{Turtle report format}\label{sec:turtle-report-format}

The first report format generated by OLIVAW is a RDF format in Turtle. It is meant to be processable and queried by machines. EARL is the vocabulary used to express the turtle reports. Two extensions have been made compared to the \href{https://www.w3.org/WAI/ER/EARL10/WD-EARL10-Guide-20120125}{EARL ontology developer guide}\footnote{\url{https://www.w3.org/WAI/ER/EARL10/WD-EARL10-Guide-20120125}}.  

The Fail class mentioned in Section \ref{sec:overview} has been extended into several subclasses in order to express some levels of severity for a given Fail outcome, namely \texttt{olivaw:MajorFail} for an outcome that would block the ontology to be deployed to production, and \texttt{olivaw:MinorFail} otherwise.

\begin{figure*}
\centering
\subfloat{
  \centering
  \includegraphics[width=.4\linewidth]{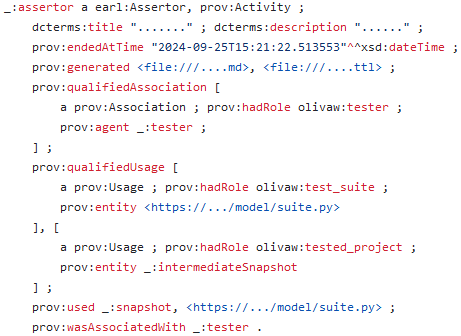}
  \label{fig:assertor-turtle}
}
\subfloat{
  \centering
  \includegraphics[width=.4\linewidth]{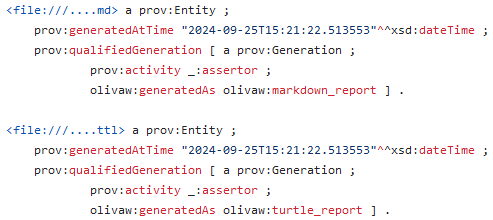}
  \label{fig:generation-turtle}
}
\hfill
\subfloat{
  \centering
  \includegraphics[width=.4\linewidth]{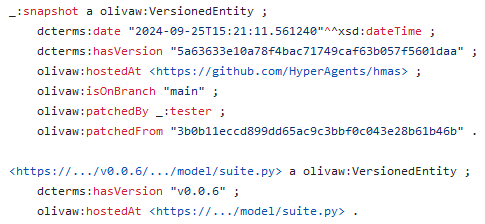}
  \label{fig:versioned-entity}
}
\subfloat{
  \centering
  \includegraphics[width=.4\linewidth]{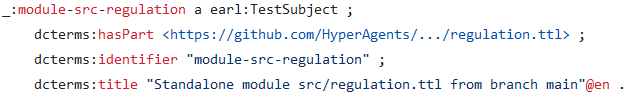}
  \label{fig:subject-turtle}
}
\hfill
\subfloat{
  \centering
  \includegraphics[width=.4\linewidth]{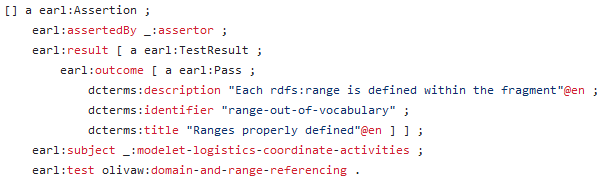}
  \label{fig:assertion-turtle}
}
\caption{Turtle reports extracts}
\label{fig:turtle-reports}
\vspace{-15pt}
\end{figure*}

The \texttt{earl:Assertor} involved in the different assertions of the test is described as an activity that involves: a person associated as developer;  the ontology used as tested subject; the executed OLIVAW test script used as test suite; both reports files (Turtle and Markdown) as activity generations.
This activity is described using the PROV ontology as shown in Figure \ref{fig:assertor-turtle}.
In this format, generated files are named by their GitHub file URIs if GitHub Actions ran the test and by their local file paths if tests were run locally. In order to qualify associations and usages, several \texttt{prov:Role} instances are defined in the OLIVAW ontology. A property has also been defined to qualify the activity generations. It is used in the report as shows Figure \ref{fig:generation-turtle}.


The test activity used two \texttt{prov:Entity} instances, namely the tested project and the test suite. These entities are of type \texttt{olivaw:VersionedEntity}. Resources of this subclass of \texttt{prov:Entity} are identified (with an \texttt{owl:hasKey} axiom) by the host URL and the entity version. When a test is run locally, it may apply to a local repository state that has no commit hash yet. In that case, the tested project version is a hash of files hashes in order to identify it from another different uncommited local state. Figure \ref{fig:versioned-entity} shows this implementation.


For each test, the \texttt{earl:TestSubject} is composed of URIs pointing to the files that were involved in the Assertion, in the origin server, as illustrated in Figure \ref{fig:subject-turtle}. 
%
%
Finally, the result is a node that can point to one to several outcomes for each Assertion. Each assertion in the report is an \texttt{earl:Assertion}. This assertion implementation can be seen in Figure \ref{fig:assertion-turtle}.


\subsection{Markdown report format}\label{sec:markdown-report-format}

The markdown report format is derived from the turtle report to provide a human-friendly version. Since this report can be very long, it includes hyperlinks to be browsed, searched and navigated easily on GitHub and to be fully integrated in that environment. The report is composed of several sections that we review next. Some examples of reports for model tests are available online
.\footnote{\url{https://github.com/Wimmics/olivaw/blob/main/docs/examples/hmas/model-test-manual-NicoRobertIn-2024-12-19T08-59-52.md}} 

The first section 
is a short explanation about the report itself and how it was obtained from the Turtle version. 
%
%
The second section 
provides information about the assertor, including the developer name, how the report generation was triggered, the triggered test suite and finally the date time of the tests. When a local test is run, the project version is identified by the hash of all the project file hashes that are not ignored. 
This allows to distinguish the head commit version from an uncommitted local version or different uncommitted local versions. In any local test report, the hash corresponding to the commit that uncommitted version is derived from will also be specified.

\begin{figure}
    \centering
    \includegraphics[width=0.8\linewidth]{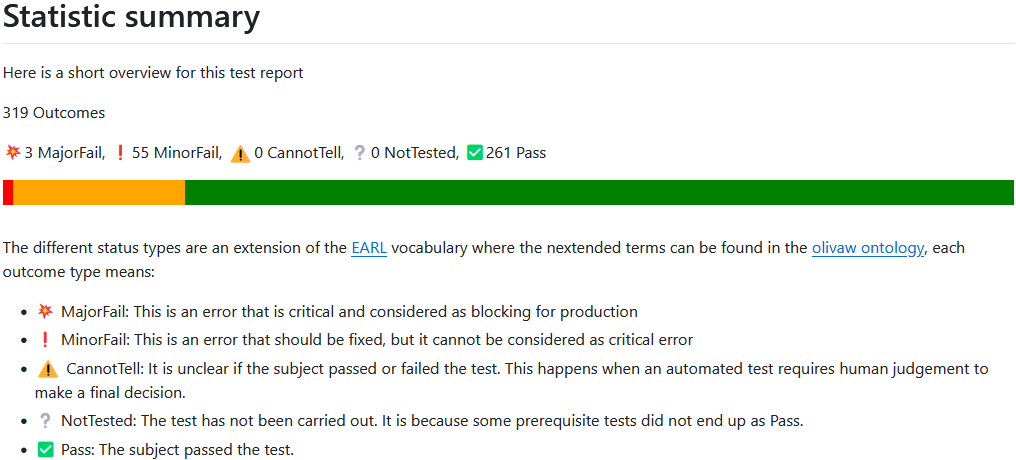}
    \caption{Statistics summary of a markdown report}
    \label{fig:statistics-summary}
    \vspace{-10pt}
\end{figure}

The third section, showed in Figure \ref{fig:statistics-summary}, is a statistic summary providing the total number of entries in the report and the number of outcomes for each type of severity an outcome can have, with a very simple bar chart and a brief explanation of the meaning of each severity.

The next sections are providing details about the outcomes for each type of outcome present in the report. Examples are provides in Section~\ref{sec:experimentation}. Each of them begins with a table that is a summary of the outcome containing the subject, the criterion that is related to that outcome, the title of the outcome and a link that leads directly to the detailed information about that given outcome. 
Each of these sections ends with an aggregation of the outcome details with, for each outcome, tables providing subject, criterion and outcome information. The subject table provides the subject natural language description and the list of the files composing the subject with their links to their GitHub related pages. The criterion table provides the criterion title and description. The outcome table provides the outcome title, description and a list of pointers. These pointers can be URIs, snippets of code or natural language messages providing useful information about the outcome. Each of these detailed outcome has a link bringing back to the related outcome summary line so the user can browse the summary, check the details of an outcome, and resume browsing again.



\section{Deployment of OLIVAW}\label{sec:deploy}

OLIVAW can be used: (i) locally as a python package providing a command line tool that can be run at development time; (ii) locally as a pre-commit connector for automatic test before commit and (iii) remotely as GitHub Composite Actions triggering on events such as push or merge for continuous validation. It can be configured as described in the rest of this section.

%
As a command line tool, OLIVAW also provides repository management commands such as for: (i) initializing a repository or a new branch; (ii) testing the project from a model, data or query perspective and (iii) flushing the reports folder. The \href{https://github.com/Wimmics/olivaw/blob/main/docs/commands.md}{OLIVAW command line documentation is available online}\footnote{\url{https://github.com/Wimmics/olivaw/blob/main/docs/commands.md}}. 
%
%
OLIVAW has a pre-commit connector usable from any repository. On trigger, it analyzes each file staged for commit and blocks commit in case it detects a blocking error. If the commit is blocked, the console outputs useful information about the files and the related blocking errors. This connector is agnostic to the way git is used (command line, desktop or anything else). The \href{https://github.com/Wimmics/olivaw/blob/main/docs/pre-commit.md}{OLIVAW pre-commit hook documentation is available online}\footnote{\url{https://github.com/Wimmics/olivaw/blob/main/docs/pre-commit.md}}.


\textit{GitHub Actions} is a continuous integration mechanism on GitHub allowing script execution on GitHub side in reaction to some repository event. A \textit{GitHub Composite Actions} is a ready-to-use \textit{GitHub Action}s script callable from a \textit{GitHub Actions} script of another repository. OLIVAW has two \textit{GitHub Composite Actions} in order to react to some actions that are made on the repository.
The first \textit{GitHub Actions} concerns the tests. It can be triggered each time a change is pushed to the repository and launches each of the desired tests. These tests can then : (a) be uploaded as GitHub artifacts in order to keep track of the history of the state of a branch; (b) be committed on the branch in order to have on GitHub the actual status of the branch and (c) update the gist files that are used to dynamically display the proper data in the \texttt{README.md} file badges.
The second \textit{GitHub Actions} assists branch creation and updates the branch badges links to the readme file when a branch is created. It also creates new gist files on GitHub to have independant badges for each branch. The \href{https://github.com/Wimmics/olivaw/blob/main/docs/actions.md}{OLIVAW Composite Actions documentation}\footnote{\url{https://github.com/Wimmics/olivaw/blob/main/docs/actions.md}} is available online.


An ontology development project can have specific requirements, including which errors are considered as blocking, which tests from the suites should not run, which files should not be tested, or a combination of these factors (e.g., which test should not run on a given file). This may be configured in the \texttt{parameters.json} file (see the \href{https://github.com/Wimmics/olivaw/blob/main/docs/parameters.md}{OLIVAW parameters file documentation} online\footnote{\url{https://github.com/Wimmics/olivaw/blob/main/docs/parameters.md}}).




\section{Experimentation and Evaluation}\label{sec:experimentation}

\subsection{First use case: the hMAS ontology project}\label{sec:hmas}
\begin{figure}
    \centering
    \vspace{-5pt}
    \includegraphics[width=0.8\linewidth]{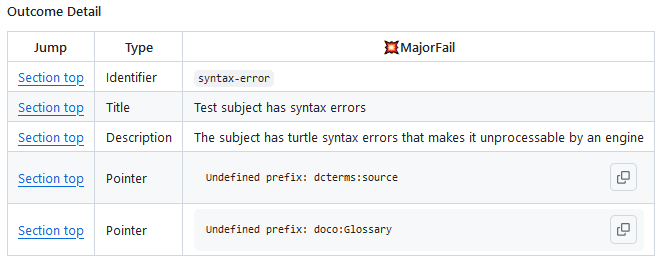}
    \caption{Example of model syntax error in markdown}
    \label{fig:model-syntax}
    \vspace{-5pt}
\end{figure}

This section illustrates the use of OLIVAW during the hMAS ontology development\footnote{\url{https://github.com/HyperAgents/hmas/}} with some examples of reports for the different types of errors we covered in the previous sections. 
The first example is the syntax error for which an example of report is shown in Figure \ref{fig:model-syntax}.

\begin{figure}
    \centering
    \vspace{-5pt}
    \includegraphics[width=0.8\linewidth]{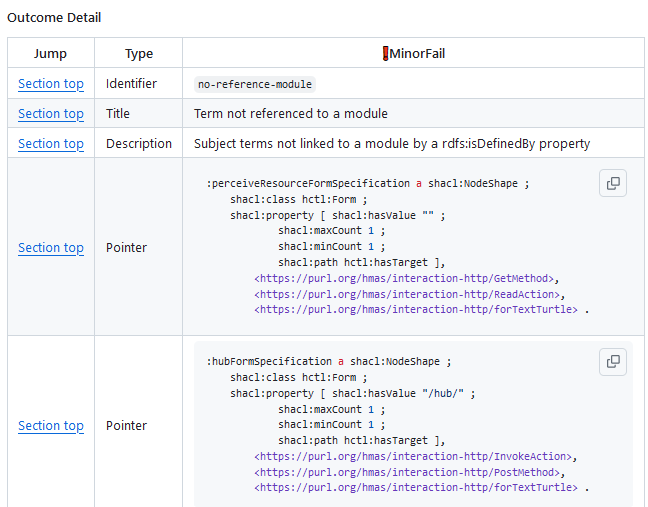}
    \caption{Example of term referencing error in markdown}
    \label{fig:model-referencing}
    \vspace{-5pt}
\end{figure}

Another error considered in hMAS is the lack of \texttt{rdfs:isDefinedBy} property on a ontology term, for which an example of report is shown in Figure~\ref{fig:model-referencing}.  

\begin{figure}
    \centering
    \vspace{-5pt}
    \includegraphics[width=0.8\linewidth]{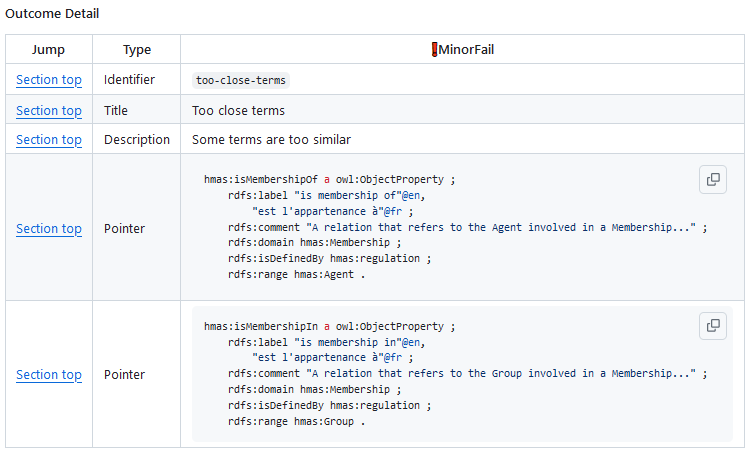}
    \caption{Example of term Differentiation error in markdown}
    \label{fig:model-Differentiation}
    \vspace{-5pt}
\end{figure}

As a last example we can highlight the excessive similarity between different terms. In hMAS the pair of terms \texttt{isMembershipOf} and \texttt{isMembershipIn} has a Levenshtein distance only equal to 2. The difference between these terms is clearly defined, but it could lead the users downstream to confusions and typos. Figure \ref{fig:model-Differentiation} shows how such an error is reported by OLIVAW.

\subsection{Generalizing to different ontologies}\label{sec:coswot}

In order to ensure the framework genericity and its usability over a large panel of ontologies, OLIVAW has also been tested against \href{https://gitlab.com/coswot/coswot-acimov/}{COSWOT}\footnote{\url{https://gitlab.com/coswot/coswot-acimov/}} and \href{https://github.com/Accord-Project/aec3po}{AEC3PO}\footnote{\url{https://github.com/Accord-Project/aec3po}} projects, which match the ACIMOV architecture, and against a few already deployed INRIA ontologies, namely \href{https://ns.inria.fr/munc}{MUNC}\footnote{\url{https://ns.inria.fr/munc}}, \href{https://ns.inria.fr/nrv}{NRV}\footnote{\url{https://ns.inria.fr/nrv}} and \href{https://ns.inria.fr/pwkso}{PWKSO}\footnote{\url{https://ns.inria.fr/pwkso}}. The reports mentioned here can be found in the \href{https://github.com/Wimmics/olivaw/tree/query/docs/examples}{examples section of OLIVAW repository}\footnote{\url{https://github.com/Wimmics/olivaw/tree/query/docs/examples}}.

Running the model test on COSWOT revealed a syntax error on the main branch. In file \texttt{core/coswot-saref.ttl} on line 112, a semi-column was forgotten, causing a parsing error on line 113. On top of this only blocking error, many outcomes related to profile diagnose have revealed that COSWOT and AEC3PO ontology are included neither in OWL EL, QL or RL profiles. The enumeration of the OWL RL incompatibility outcomes indicated that, in order to be OWL RL compatible, the ontologies should avoid the use of \texttt{owl:disjointUnionOf} and avoid the use of \texttt{owl:Restriction} as objects of the predicates \texttt{rdfs:subClassOf}, \texttt{rdfs:domain} and \texttt{rdfs:range}. If these changes are made on the ontology, it would make them OWL RL compatible.

The different  INRIA ontologies tested are also really close to comply with the OWL RL profile. MUNC should avoid using \texttt{owl:ReflexiveProperty}. NRV is using \texttt{owl:unionOf} but not as a subclass expression and uses \texttt{owl:disjointUnionOf}. PWKSO uses restrictions as objects of \texttt{rdfs:subClassOf}.
Once these incompatibility issues are highlighted by OLIVAW, it is then up to the developers to decide if these axioms are needed for the description of the domain, or if the removal of these axioms is a fair price to pay for being in a simpler OWL profile.



\section{Conclusion}\label{sec:conclusion}

In this article, we presented OLIVAW (Ontology Long-lived Integration Via ACIMOV Workflow), a tool that implements the ACIMOV methodology on GitHub relying on W3C Standards to assist the lifecycle of modularised and multi-domain ontologies.
We introduced the general principles and each family of tests it can perform and the reports it produces.
We explained different alternatives to use OLIVAW and we reported on our experimentation and evaluations against different ontologies of various sizes and application domains. 
OLIVAW is available on GitHub under the LGPL-2.1 license, with its documentation and examples in \href{https://github.com/Wimmics/olivaw}{OLIVAW GitHub repository}\footnote{\url{https://github.com/Wimmics/olivaw}}.

There are many other features that we are considering for OLIVAW. First to create GitLab CI/CD implementations of the existing GitHub Actions in order for OLIVAW to be fully compatible with GitLab. Secondly, to add some ontology integration control features in order to address large-scale projects. And finally to provide some deployment features and automatic tools that would help developers at other development steps than the testing step.

\bibliographystyle{IEEEtran}
\bibliography{references}

\begin{thebibliography}{10}
\providecommand{\url}[1]{#1}
\csname url@samestyle\endcsname
\providecommand{\newblock}{\relax}
\providecommand{\bibinfo}[2]{#2}
\providecommand{\BIBentrySTDinterwordspacing}{\spaceskip=0pt\relax}
\providecommand{\BIBentryALTinterwordstretchfactor}{4}
\providecommand{\BIBentryALTinterwordspacing}{\spaceskip=\fontdimen2\font plus
\BIBentryALTinterwordstretchfactor\fontdimen3\font minus
  \fontdimen4\font\relax}
\providecommand{\BIBforeignlanguage}[2]{{%
\expandafter\ifx\csname l@#1\endcsname\relax
\typeout{** WARNING: IEEEtran.bst: No hyphenation pattern has been}%
\typeout{** loaded for the language `#1'. Using the pattern for}%
\typeout{** the default language instead.}%
\else
\language=\csname l@#1\endcsname
\fi
#2}}
\providecommand{\BIBdecl}{\relax}
\BIBdecl

\bibitem{hannou:hal-04187236}
\BIBentryALTinterwordspacing
F.-Z. Hannou, V.~Charpenay, M.~Lefran{\c c}ois, C.~Roussey, A.~Zimmermann, and
  F.~Gandon, ``{The ACIMOV Methodology: Agile and Continuous Integration for
  Modular Ontologies and Vocabularies},'' in \emph{{MK 2023 - 2nd Workshop on
  Modular Knowledge associated with FOIS 2023 - the 13th International
  Conference on Formal Ontology in Information Systems}}, Sherbrooke, Canada,
  Jul. 2023. [Online]. Available: \url{https://laas.hal.science/hal-04187236}
\BIBentrySTDinterwordspacing

\bibitem{abdelghany2019agile}
A.~S. Abdelghany, N.~R. Darwish, and H.~A. Hefni, ``An agile methodology for
  ontology development,'' \emph{International Journal of Intelligent
  Engineering and Systems}, vol.~12, no.~2, pp. 170--181, 2019.

\bibitem{takhom2020collaborative}
A.~Takhom, S.~Usanavasin, T.~Supnithi, and P.~Boonkwan, ``A collaborative
  framework supporting ontology development based on agile and scrum model,''
  \emph{IEICE TRANSACTIONS on Information and Systems}, vol. 103, no.~12, pp.
  2568--2577, 2020.

\bibitem{conf/fomi/ShariflooS08}
A.~A. Sharifloo and M.~Shamsfard, ``Using agility in ontology construction.''
  in \emph{Formal Ontologies Meet Industry}, ser. Frontiers in Artificial
  Intelligence and Applications, vol. 174.\hskip 1em plus 0.5em minus
  0.4em\relax IOS Press, 2008, pp. 109--119.

\bibitem{hristozova2002extreme}
M.~Hristozova and L.~Sterling, ``An extreme method for developing lightweight
  ontologies,'' in \emph{In Workshop on Ontologies in Agent Systems, 1st
  International Joint Conference on Autonomous Agents and Multi-Agent
  Systems}.\hskip 1em plus 0.5em minus 0.4em\relax Citeseer, 2002.

\bibitem{de2013organizing}
R.~de~Almeida~Falbo, M.~P. Barcellos, J.~C. Nardi, and G.~Guizzardi,
  ``Organizing ontology design patterns as ontology pattern languages,'' in
  \emph{The Semantic Web: Semantics and Big Data: 10th International
  Conference, ESWC 2013, Montpellier, France, May 26-30, 2013. Proceedings
  10}.\hskip 1em plus 0.5em minus 0.4em\relax Springer, 2013, pp. 61--75.

\bibitem{cummings2018lean}
J.~Cummings and D.~Stacey, ``Lean ontology development: An ontology development
  paradigm based on continuous innovation.'' in \emph{Knowledge Engineering and
  Ontology Development}, 2018, pp. 365--372.

\bibitem{peroni2016samod}
S.~Peroni, ``Samod: an agile methodology for the development of ontologies,''
  in \emph{OWL: Experiences and Directions Workshop, OWL reasoner evaluation
  workshop, OWLED-ORE 2016}, 2016, pp. 1--14.

\bibitem{uschold_gruninger_1996}
M.~Uschold and M.~Gruninger, ``Ontologies: principles, methods and
  applications,'' \emph{The Knowledge Engineering Review}, vol.~11, no.~2, p.
  93–136, 1996.

\bibitem{ciroku2023supporting}
F.~Ciroku, ``Supporting requirement elicitation and ontology testing in
  knowledge graph engineering,'' 2023.

\bibitem{halilaj2016vocol}
L.~Halilaj, N.~Petersen, I.~Grangel-Gonz{\'a}lez, C.~Lange, S.~Auer, G.~Coskun,
  and S.~Lohmann, ``Vocol: An integrated environment to support
  version-controlled vocabulary development,'' in \emph{European Knowledge
  Acquisition Workshop}.\hskip 1em plus 0.5em minus 0.4em\relax Springer, 2016,
  pp. 303--319.

\bibitem{alobaid2019automating}
A.~Alobaid, D.~Garijo, M.~Poveda-Villal{\'o}n, I.~Santana-Perez,
  A.~Fern{\'a}ndez-Izquierdo, and O.~Corcho, ``Automating ontology engineering
  support activities with ontoology,'' \emph{Journal of Web Semantics},
  vol.~57, p. 100472, 2019.

\bibitem{matentzoglunicolas}
\BIBentryALTinterwordspacing
N.~Matentzoglu, C.~Mungall, and D.~Goutte-Gattat, ``Ontology development kit,''
  Jul. 2021. [Online]. Available: \url{https://doi.org/10.5281/zenodo.6257507}
\BIBentrySTDinterwordspacing

\bibitem{jackson2019robot}
R.~C. Jackson, J.~P. Balhoff, E.~Douglass, N.~L. Harris, C.~J. Mungall, and
  J.~A. Overton, ``Robot: a tool for automating ontology workflows,'' \emph{BMC
  bioinformatics}, vol.~20, no.~1, pp. 1--10, 2019.

\bibitem{neuhaus2022infrastructure}
D.~Allemang, P.~Garbacz, P.~Gradzki, E.~Kendall, and R.~Trypuz, ``An
  infrastructure for collaborative ontology development, lessons learned from
  developing the financial industry business ontology {(FIBO)},'' in
  \emph{Formal Ontology in Information Systems}.\hskip 1em plus 0.5em minus
  0.4em\relax IOS Press, 2022.

\bibitem{bader2020international}
S.~Bader, J.~Pullmann, C.~Mader, S.~Tramp, C.~Quix, A.~W. M{\"u}ller,
  H.~Aky{\"u}rek, M.~B{\"o}ckmann, B.~T. Imbusch, J.~Lipp \emph{et~al.}, ``The
  international data spaces information model--an ontology for sovereign
  exchange of digital content,'' in \emph{ISWC 2020, International Semantic Web
  Conference, Athens, November 2--6}.\hskip 1em plus 0.5em minus 0.4em\relax
  Springer, 2020, pp. 176--192.

\bibitem{saref10154112}
R.~Garc{\'{\i}}a{-}Castro, M.~Lefran{\c{c}}ois, M.~Poveda-Villal{\'o}n, and
  L.~Daniele, \emph{The ETSI SAREF Ontology for Smart Applications: A Long Path
  of Development and Evolution}.\hskip 1em plus 0.5em minus 0.4em\relax
  Wiley-IEEE Press, 2023, pp. 183--215.

\bibitem{lefranccois2023saref}
M.~Lefran{\c{c}}ois and D.~Gnabasik, ``The saref pipeline and portal—an
  ontology verification framework,'' in \emph{International Semantic Web
  Conference}.\hskip 1em plus 0.5em minus 0.4em\relax Springer, 2023, pp.
  134--151.

\bibitem{publio2022ontolo}
G.~C. Publio, J.~E.~L. Gayo, G.~F. Colunga, and P.~Menend{\'e}z, ``Ontolo-ci:
  Continuous data validation with shex,'' in \emph{Proceedings of Poster and
  Demo Track and Workshop Track of the 18th International Conference on
  Semantic Systems co-located with 18th International Conference on Semantic
  Systems ({SEMANTiCS} 2022)}, 2022.

\bibitem{Lanthaler:14:RCA}
M.~Lanthaler, D.~Wood, and R.~Cyganiak, ``{RDF} 1.1 concepts and abstract
  syntax,'' W3C, {W3C} Recommendation, Feb. 2014,
  https://www.w3.org/TR/2014/REC-rdf11-concepts-20140225/.

\bibitem{Parsia:12:OWO}
B.~Parsia, P.~Patel-Schneider, S.~Rudolph, P.~Hitzler, and M.~Kr{\"{o}}tzsch,
  ``{OWL} 2 web ontology language primer (second edition),'' W3C, {W3C}
  Recommendation, Dec. 2012,
  https://www.w3.org/TR/2012/REC-owl2-primer-20121211/.

\bibitem{Seaborne:13:SQL}
A.~Seaborne and S.~Harris, ``{SPARQL} 1.1 query language,'' W3C, {W3C}
  Recommendation, Mar. 2013,
  https://www.w3.org/TR/2013/REC-sparql11-query-20130321/.

\bibitem{Kontokostas:17:SCL}
D.~Kontokostas and H.~Knublauch, ``Shapes constraint language ({SHACL}),'' W3C,
  {W3C} Recommendation, Jul. 2017,
  https://www.w3.org/TR/2017/REC-shacl-20170720/.

\bibitem{Sahoo:13:PTP}
S.~Sahoo, T.~Lebo, and D.~McGuinness, ``{PROV}-o: The {PROV} ontology,'' W3C,
  {W3C} Recommendation, Apr. 2013,
  https://www.w3.org/TR/2013/REC-prov-o-20130430/.

\bibitem{Abou-Zahra:17:ERL}
S.~Abou-Zahra, ``Evaluation and report language ({EARL}) 1.0 schema,'' W3C,
  {W3C} Note, Feb. 2017,
  https://www.w3.org/TR/2017/NOTE-EARL10-Schema-20170202/.

\bibitem{ceres:hal-04170333v1}
\BIBentryALTinterwordspacing
R.~C{\'e}r{\`e}s, O.~Corby, E.~Demairy, and F.~Gandon, ``{Corese},'' Jul. 2023.
  [Online]. Available: \url{https://hal.science/hal-04170333}
\BIBentrySTDinterwordspacing

\bibitem{rfc3986}
\BIBentryALTinterwordspacing
T.~Berners-Lee, R.~T. Fielding, and L.~M. Masinter, ``{Uniform Resource
  Identifier (URI): Generic Syntax},'' RFC 3986, Jan. 2005. [Online].
  Available: \url{https://www.rfc-editor.org/info/rfc3986}
\BIBentrySTDinterwordspacing

\end{thebibliography}

\end{document}